\documentclass[referee]{aa}
\usepackage{latexsym}
\usepackage{amsmath}
\usepackage{amssymb}
\usepackage{graphicx}
\usepackage[english]{babel}
\title{Evolution of deuterium, $^{3}$He and $^{4}$He in the Galaxy}
\author{C. Chiappini\inst{1}\and A. Renda\inst{2}\and F. Matteucci\inst{2,1}}
\institute{Osservatorio Astronomico di Trieste, Via G. B. Tiepolo 11,\newline I - 34131 Trieste, Italia \and Dipartimento di Astronomia, Universita' degli Studi di Trieste,\newline Via G. B. Tiepolo 11, I - 34131 Trieste, Italia}
\date{Received / Accepted}
\abstract{In this work we present the predictions of the ``two-infall model'' concerning the evolution of D, $^{3}$He and $^{4}$He in the solar vicinity, as well as their distribution along the Galactic disk. Our results show that, when adopting detailed yields taking into account the extra-mixing process in low and intermediate mass stars, the problem of the overproduction of $^{3}$He by the chemical evolution models is solved. The predicted distribution of $^{3}$He along the disk is also in agreement with the observations. We also predict
the distributions of D/H, D/O and D/N along the disk, in particular
D abundances close to the primordial value are predicted in the outer regions of the Galaxy. The predicted D/H, D/O and D/N abundances in the local interstellar medium are in agreement with the mean values observed by the Far Ultraviolet Spectroscopic Explorer mission, although a large spread in the D abundance 
is present in the data.
Finally, by means of our chemical evolution model, we can constrain the primordial value of the deuterium abundance, and we find a value of (D/H)$_{\rm p} \lesssim$ 4 $\cdot$ 10$^{-5}$ which implies $\Omega_{b}h^2 \gtrsim $ 0.017, in agreement with the values from the Cosmic Microwave Background radiation analysis. This value in turn implies a primordial $^4$He abundance Y$_{\rm p} \gtrsim $ 0.244.

}
\begin{document}
\maketitle
\section{Introduction}
Chemical evolution models are useful tools to derive the primordial abundances 
of light elements, such as D, $^{3}$He and $^{4}$He, and to give informations on stellar 
nucleosynthesis. 
It is well known that while the abundances of metals increase in time with galactic 
evolution, 
the abundance of D is always decreasing since this element is only destroyed in stellar 
nucleosynthesis (for another view see Mullan \& Linsky 1999, who suggest possible D 
ejection from flares into the stellar wind). On the other hand, $^{3}$He and $^{4}$He 
should be partially produced, and partially burned into heavier elements.  
\newline
In this paper we show the predictions of the ``two-infall model'' (Chiappini et al.  2001), 
concerning the evolution of D, $^{3}$He and $^{4}$He in the solar vicinity and their distribution along the galactic disk.
\newline
This model assumes two main infall episodes for the formation of the halo and part of 
the thick disk, and the thin disk, respectively. The timescale for the formation of the 
thin disk is much longer than that of the halo, implying that the infalling gas forming 
the thin disk comes not only from the halo but mainly from the intergalactic medium. 
The timescale for the formation of the thin disk is assumed to be a function of the 
galactocentric distance, leading to an inside-out picture for the Galaxy disk formation.
\newline
The two-infall model differs from other models in the literature mainly in two aspects: it considers an almost independent evolution between the halo and thin disk components (see also Pagel \& Tautvaisiene  1995), and it assumes a threshold in the star formation process (see van der Hulst et al. 1993; Kennicutt 1989, 1998; Martin \& Kennicutt 2001).
The model well reproduces the majority of observational constraints about the abundances of heavy elements both locally and in the whole disk (Chiappini et al. 1997, 2001).
\newline
The main novelty of this paper concerns the $^{3}$He nucleosynthesis prescriptions. Starting from the paper of Iben (1967), stellar models have predicted that low-mass stars are strong producers of $^{3}$He. However, when stellar production of $^{3}$He
is included in models for the chemical evolution of the Galaxy, no agreement between observed and predicted $^{3}$He abundances can be found. In particular, the models predict an overproduction of $^{3}$He (see Rood et al. 1976; Dearborn et al. 1996; 
Prantzos 1996; Chiappini et al. 1997). 
\newline
On the other hand, there is evidence for $^{3}$He production 
in stars from the observations of this isotope in 
Galactic Planetary nebulae, such as NGC 3242, (Balser et al. 1997; Balser et al. 1999), 
and also from the observations of the $^{12}$C/$^{13}$C ratio, 
which is related to the possible occurrence of ``extra-mixing'', 
with consequent $^{3}$He destruction, in low and intermediate mass stars,
the main producers of $^{3}$He in the universe. 
The extra-mixing process has been suggested to solve the $^{3}$He overproduction problem.
Recent stellar models suggested that 
thermal instabilities might occur during the red giant branch (RGB) phase in low mass 
stars and during the early asymptotic giant branch (AGB) 
phase in intermediate mass stars, leading to a mixing between $^{3}$He-enriched material 
from the external envelope and the material which suffer H-shell burning. 
These processes should induce $^{3}$He destruction in favor of $^{7}$Li production 
(Sackmann \& Boothroyd 1999a; Palacios et al. 2001). 
Extra-mixing should also decrease the $^{12}$C/$^{13}$C ratio
(see Sect. 3 for a detailed description). 
This justifies the low observed $^{12}$C/$^{13}$C ratio in RGB stars and in Li-rich giants. 
\newline
However, not every star in the low and intermediate mass range should suffer from extra-mixing. 
For example, recent Hubble Space Telescope observations of NGC 3242, (see Palla et al. 
2002), suggests a lower limit $^{12}$C/$^{13}$C $>$ 38, 
in agreement with standard stellar models. 
Therefore, the lack of the $^{13}$C line and the presence of the $^{3}$He line in 
the spectrum of this planetary nebula suggest that the progenitor star did not undergo a 
phase of deep extra-mixing during the 
latest stages of its evolution. Charbonnel \& do Nascimento (1998) 
collected the observations of the carbon isotopic ratio in field and 
cluster giant stars, and their conclusion was that 93$\%$ of evolved 
stars undergo the extra-mixing on the Red Giant Branch, and are 
expected to destroy, at least partially, their $^{3}$He.    
Extra-mixing along the RGB is supported by recent 
observations of metal-poor field stars (Gratton et al. 2000) and of stars along the 
RGB phase in the thick disk of the Galaxy (Tautvaisiene et al. 2001). 
\newline
In the present work we take into account the detailed prescriptions 
about $^{3}$He production/destruction by Sackmann \& Boothroyd (1999a, b) and their
dependence upon stellar 
mass and metallicity. The $^{3}$He production is strictly correlated with 
the destruction of D and the production of $^{4}$He. Therefore, we consider all these elements.
We compare the predictions for the time-evolution of the abundances of D, 
$^{3}$He and $^{4}$He
and for their distribution along the disk, with the available data. 
The latest data on $^{3}$He/H, which are observations of the hyperfine 
transition of $^{3}$He$^{+}$ at 8.665 GHz (3.46cm), are from Bania et al. (2002), 
and show a flat distribution around a value of about 1.5$\cdot~10^{-5}$  
(with the exception of the observation of a larger $^{3}$He abundance in a few nebulae such as NGC 3242). 
The latest published data for D/H, D/O, D/N, are from the Far Ultraviolet Spectroscopic 
Explorer mission (Moos et al. 2002) and refer to the local interstellar medium 
(LISM). These observations (Jenkins et al. 1999; Sonneborn et al. 2000; 
Friedman et al. 2002) show a large dispersion in the deuterium abundances beyond 
a few tenth of kpc in the solar vicinity 
which is not easy to justify since these effects are not seen in the oxygen and nitrogen abundances. 
\newline
The paper is organized as follows: in Sect. 2 we describe the chemical evolution model, 
in Sect. 3 we discuss in detail the adopted nucleosynthesis prescriptions for 
D, $^{3}$He and $^{4}$He. In Sect. 4 the results are presented and discussed and finally, 
in Sect. 5 some conclusions are drawn.

\section{The model}
The model assumes that the Galaxy forms out of two main accretion episodes almost completely 
disentangled. During the first episode, the primordial gas collapses very quickly and forms 
the spheroidal components, halo and bulge; during the second episode, the thin-disk forms, 
mainly by accretion of a significant fraction of matter of primordial chemical composition 
plus traces of halo gas. The disk is built-up in the framework of the inside-out scenario of Galaxy formation, which ensures the formation of abundance gradients along the disk
(Larson 1976; Matteucci \& Fran\c cois 1989). As in Matteucci \& Fran\c cois (1989), 
the Galactic disk is approximated by several independent rings, 2 kpc wide, without exchange of matter between them. The rate of accretion of matter in each shell is:
\begin{equation}
\frac{d\Sigma_I(R, t)}{dt} = A(R)\,e^{- t/\tau_{H}} + B(R)\,e^{- (t - t_{max})/\tau_{D}},
\end{equation}
where $\Sigma_I(R, t)$ is the surface mass density of the infalling material, 
which is assumed to have primordial chemical composition; $t_{max}$ is the time 
of maximum gas accretion onto the disk, coincident with the end of the halo/thick-disk phase and set here equal to 1 Gyr; $\tau_{H}$ and $\tau_{D}$ are the timescales for mass accretion onto the halo/thick-disk and thin-disk components, respectively. In particular, $\tau_{H}$ = 0.8 Gyr and, according to the inside-out scenario, $\tau_{D}(R)$ = 1.03$(R$/kpc)$-$1.27 Gyr
(see Chiappini et al. 2001). 
A(R) and B(R) are derived by the condition of reproducing the present
total surface mass density distribution in the solar vicinity (see
Matteucci 2001).
\newline
The SFR adopted here has the same formulation as in Chiappini et al. (1997): 
\begin{equation}
\psi(R, t) = \nu(t)\Big( \frac{\Sigma(R, t)}{\Sigma(R_{\odot}, t)} \Big) ^{2(k - 1)}\Big( \frac{\Sigma(R, t_{Gal})}{\Sigma(R, t)} \Big)^{k - 1}G_{gas}^{k}(R, t),
\end{equation}
where $\nu(t)$ is the efficiency of the star formation process, $\Sigma(R,t)$ is 
the total mass surface density at a given radius $R$ and given time $t$, $\Sigma(R_{\odot}, t)$ 
is the total mass surface density at the solar position, $G_{gas}(R, t)$ is the 
gas surface density normalized to the present time total surface mass density in the disk $\Sigma_{D}(R,t_{Gal})$, 
t$_{Gal}$ = 14 Gyr is the age of the Galaxy, and R$_{\odot}$ = 8 kpc is the solar 
galactocentric distance (see Reid 1993). The gas surface density exponent is $k$ = 1.5, 
in order to ensure a good fit to the observational constraints in 
the solar vicinity. This value is also in agreement with the observational results of 
Kennicutt (1998), and with N-body simulation results by Gerritsen \& Icke (1997).
\newline
The efficiency of star formation is set to $\nu$ = 2 Gyr$^{-1}$, for the Galactic halo, 
whereas it is $\nu$ = 1 Gyr$^{-1}$ for the disk; this to ensure the best fit to the 
observational features in the solar vicinity. The efficiency $\nu$ becomes zero when 
the gas surface density drops below a certain critical threshold. We adopt a threshold 
density $(\sigma_{gas})_{th}$ $\simeq$ 4 $M_{\odot}$ pc$^{-2}$ for the Galactic halo, whereas it is 
$(\sigma_{gas})_{th}$ $\simeq$ 7 $M_{\odot}$ pc$^{-2}$ for the disk (see Chiappini et al. 2001). 
\newline
The IMF is that of Scalo (1986), assumed to be constant in time and space (for the
effects of a variable IMF, see Chiappini et al. 2000). 
\newline
The yields of metals are from van den Hoek \& Groenewegen (1997) 
for the low and intermediate 
mass stars; from Thielemann et al. (1996) for the nucleosynthesis from SNeII; 
from Thielemann et al. (1993) for the nucleosynthesis from SNeIa; 
from Jos\'e \& Hernanz (1998) for the nucleosynthesis from nova outbursts
for the elements $^{7}$Li, $^{13}$C and $^{15}$N. See Sect. 3
for the detailed nucleosynthesis prescriptions of D, $^{3}$He and $^{4}$He.
\newline
The present time surface mass density distribution of the disk is exponential with scale 
length $R_{D}$ = 3.5 kpc normalized to 
$\Sigma_{D}(R_{\odot}, t_{Gal})$ = 54 $M_{\odot}$ pc$^{-2}$ (see Romano et al. 2001). 
The main parameters are presented in Table 1. 
\begin{table}
\caption{Model parameters common to all the models.}
\begin{tabular}{ll}
\hline
&\\
$\tau_{H}$ & 0.8 Gyr, timescale for mass accretion\\
           & onto the halo/thick-disk\\
$\tau_{D}(R)$ & 1.03$(R$/kpc)$-$1.27 Gyr, timescale for mass accretion\\
              & onto the thin-disk\\
$t_{max}$ & 1 Gyr, the time of maximum gas accretion\\
          & onto the thin-disk\\
$\nu(t)$ & 2 Gyr$^{-1}$, the efficiency of star formation,\\
         & for the halo/thick-disk\\
$\nu(t)$ & 1 Gyr$^{-1}$, the efficiency of star formation,\\
         & for the thin-disk\\
$(\sigma_{gas})_{threshold}$ & $\simeq$ 4 $M_{\odot}$ pc$^{-2}$, the gas surface density threshold\\
                     & below which $\nu$ = 0 for the halo/thick-disk\\
$(\sigma_{gas})_{threshold}$ & $\simeq$ 7 $M_{\odot}$ pc$^{-2}$, the gas surface density threshold\\
                     & below which $\nu$ = 0 for the thin-disk\\
$k_{Schmidt~law}$ & 1.5\\
$x_{IMF}$ & 1.35, if M $<$ 2 $M_{\odot}$\\ 
$x_{IMF}$ &  1.7, if 2 $M_{\odot}<$ M $<$ 80 $M_{\odot}$\\ 
$R_{D}$ & 3.5 kpc, scale length for the exponential distribution\\
        & of the disk present surface mass density\\
$\Sigma_{D}(R_{\odot},t_{Gal})$ & 54 $M_{\odot}$ pc$^{-2}$, the present surface mass density of the disk\\
$R_{g,\odot}$ & 8 kpc, solar galactocentric distance\\
$t_{Gal}$ & 14 Gyr, age of the Galaxy\\
&\\
\hline
\end{tabular}
\label{table1}
\end{table}
\begin{table}
\caption{Baryon to photon ratio $\eta$ and $\Omega_{b}h^{2}$, using the observed value of D/H in high-redshift objects as baryometer, and Standard Big Bang Nucleosynthesis. 
\newline
For a comparison, $\Omega_{b}h^{2}_{CMB} = 0.032^{+0.009}_{-0.008}$, at 95\% Confidence Level, see Jaffe et al. (2001).}
\begin{tabular}{lll}
& O'Meara et al. (2001) & Levshakov et al. (2002)\\
\hline
&&\\
$10^{-5}\cdot$(D/H) & $3.0\pm 0.4$ & $3.75\pm 0.25$\\
$10^{-10}\cdot\eta$ & $5.6\pm 0.5$ & 4.37 - 5.32\\ 
$\Omega_{b}h^{2}$ & $0.0205\pm 0.0018$ & 0.016$<\Omega_{b}h^{2}<$0.020\\
&&\\
\hline
\end{tabular}
\label{table2}
\end{table}

\subsection{The adopted primordial abundances} 

In Table 2 we show some of the most recent estimates of the primordial D abundance. 
These are mainly obtained from observations of Damped Lyman Alpha (DLA) and Lyman limit systems, which are supposed to be objects with N(HI)$>$2$\cdot10^{20}$~atoms~cm$^{-2}$ and N(HI)$>$3$\cdot10^{17}$~atoms~cm$^{-2}$ respectively, and where the D abundance is probably 
close to the primordial value. In Table 2 we also show the cosmological parameters 
derived using D as a baryometer from Standard Big Bang Nucleosynthesis.
Within the Standard Big Bang scenario the deuterium abundance is a monotonic,
rapidly decreasing function of the baryon density $\eta$. Once the Big Bang 
Nucleosynthesis begins, D is rapidly burned into $^3$He and $^4$He. Moreover, the
higher the baryon ratio, the faster the burning and the less D survives (Steigman
2000). This strong baryon abundance dependence combined with the simplicity of 
the evolution of D/H in time (being only destroyed inside stars) is responsible 
for the unique role of deuterium as a baryometer. 
Given the above, we will consider the different observational
estimates of the primordial D/H abundance and, for each of these measurements,
obtain the primordial values of $^3$He/H and $^4$He/H by comparison with the 
theoretical predictions of Burles et al. (1999).
\newline
In particular, O'Meara et al. (2001) take the weighted 
mean of D/H values from absorbers 
toward QSOs, PKS 1937-1009, Q1009+2956, HS0105+1619, and find 
(D/H)$=$3.0 $\pm 0.4\cdot 10^{-5}$, whereas Pettini \& Bowen (2001) take the weighted 
mean of the D/H values from absorbers toward QSOs, HS0105+1619, 
Q2206-199, Q0347-3819 and find (D/H)$=$ 2.2 $\pm 0.2\cdot 10^{-5} $. 
D'Odorico et al. (2001) find (D/H)$=$ 2.24$\pm 0.67\cdot 10^{-5}$ 
in the absorber toward QSO 0347-3819; 
the latest analysis of this object by Levshakov at al. (2002) gives 
(D/H)$=$3.75 $\pm 0.25\cdot 10^{-5} $. 
We have also run models with a 
high initial abundance of deuterium (D/H $\approx 10^{-4}$) as suggested
by earlier observations of absorbers 
(for example Songaila et al. 1994; Carswell et al. 1994; 
Rugers \& Hogan 1996; Burles et al. 1999; Tytler et al. 1999). 
These observations suggest a high primordial deuterium, but are uncertain, 
and can be compromised by contaminating H (see Kirkman et al. 2001). 
Moreover, models that can reproduce the bulk of 
the observational data in the Galaxy
predict only a modest D destruction (see also Tosi et al. 1998),
thus implying a low primordial D abundance.
\newline
The observations of the Proto Solar Cloud (PSC) 
(Geiss \& Gloeckler 1998), and of the 
LISM (Linsky 1998 and Moos et al. 2002), also represent 
important constraints to the deuterium initial abundance, in the sense that the primordial D abundance should be higher than the D abundance in the PSC and this in turn should be higher than in the LISM.
\newline
In Table 3 we show the initial abundances of D, $^{3}$He and $^{4}$He adopted in our models. We note that
the primordial abundance of $^{3}$He/H obtained in this way for cases II to IV is
within the range estimated by Bania et al. (2002). These authors obtained
a value of $^{3}$He/H = 1.1 $\pm$ 0.2 $\times$ 10$^5$ for a HII region located
at 16.9 kpc from the Galactic center. They suggested this value to be close to the
primordial one given to the fact that in the outer parts of the disk 
less contamination by stellar activity is expected. Note that
the primordial values of $^{3}$He/H obtained
when assuming a high primordial D are not in good agreement with the value
suggested by Bania et al. (2002). For $^{4}$He the values shown in Table 3 can
be compared with the recent values suggested by Izotov \& Thuan (1998),
Viegas et al. (2000) and Gruenwald et al. (2002). It should be noted that
the primordial $Y_P$ estimates still suffer from important systematic errors (see Pagel 2000).
The primordial abundances of model I are from Chiappini \& Matteucci (2000).

\section{Nucleosynthesis prescriptions for the light elements}
\subsection{Deuterium}
In the present models, our choice is to assume that deuterium is only destroyed by stellar nucleosynthesis (see for example Mazzitelli \& Moretti 1980).
\subsection{$^{3}$He}
$^{3}$He is produced in low and intermediate mass stars during H-burning, but is also destroyed to form heavier species. Besides the normal nucleosynthesis processes, other mechanisms can destroy $^{3}$He.
\newline
In particular, extra-mixing could destroy a significant amount of $^{3}$He in low and intermediate 
mass stars, although the physics of the driving mechanism is not yet firmly established (see below).
\subsubsection{Extra-mixing in low and intermediate mass stars}
$^{3}$He burning could be affected by extra-mixing processes in low and intermediate mass stars. In 
fact, extra-mixing is probably related both to the mechanism which leads to the formation of Li-rich giants, and to the mechanism which could modify the $^{12}$C/$^{13}$C ratio, in the same mass range. In standard models, during the first dredge-up, which follows the exhaustion of the hydrogen burning in the core, the convective envelope reaches the region with freshly synthesized $^{3}$He. In the absence of extra-mixing processes, the $^{3}$He in the envelope survives in the star until it is ejected into the ISM by stellar winds and planetary nebulae. 
\newline
In particular, in low-mass stars, which ascend the RGB with a degenerate He core, the extra-mixing process could help to connect the $^{3}$He enriched envelope to the hotter inner region near the H-burning shell; $^{3}$He could be then burned via the Cameron \& Fowler (1971) mechanism, to create $^{7}$Li via $^{3}He(\alpha,\gamma)^{7}Be$, which quickly decays into $^{7}$Li via $^{7}Be(e^{-},\nu_{e})^{7}Li$. For the freshly synthesized $^{7}$Li to survive, the $^{7}$Be must be transported rapidly to cooler regions before the decay occurs. Depending on the mixing mechanism and its efficiency, which may vary from star to star, the $^{7}$Li may or may not reach the stellar surface. Models with extra-mixing (see for example Charbonnel 1995; Charbonnel et al. 1998, 2000) show a $^{12}$C/$^{13}$C ratio which is below the standard value.
\newline
The evolutionary track of an intermediate-mass star is different because of the faster RGB phase, which inhibits the occurrence of extra-mixing on this short period. However, extra-mixing could occur, again via a Cameron \& Fowler (1971) like process, during the early AGB phase, after the exhaustion of helium burning in the core, followed by the He-shell burning.
\newline
In stars with initial mass larger than $\simeq$ 4 M$_{\odot}$, where the shift of helium burning from the core to a shell causes the extinction of the hydrogen shell, deep-mixing could  
also be inhibited, but hot-bottom burning\footnote{In this case the convective envelope reaches the top of the H-burning shell, so that nucleosynthesis occurs at the bottom of the convective envelope.} could again lead to $^{7}$Li production (see Sackmann \& Boothroyd 1999a; Charbonnel \& Balachandran 2000; Palacios et al. 2001, for further discussion and details).
\newline
Sackmann \& Bothroyd (1999a) proposed a model, where deep extra-mixing transports envelope material into the outer wing of the hydrogen burning shell, where it undergoes partial nuclear processing, and then transports the material back out to the envelope. 
\newline
Charbonnel \& Balachandran (2000) pointed out that in this way Li-rich giants would be produced all along the RGB, whereas their data suggest $^{7}$Li production in two distinct episodes, at the luminosity bump phase of the low mass stars\footnote{The RGB luminosity function ``bump'' phase should be caused by a slower rate of evolution for stars in this phase, because of the contact between the H-burning shell and the H-rich, previously mixed, zone (Charbonnel et al. 1998, 2000).} and in the early AGB phase of low and intermediate mass stars. 
\newline
Palacios et al. (2001), showed that under certain conditions a thermal instability can occur, triggered by the nuclear burning of the freshly synthesized $^{7}$Li in a $^{7}$Li burning shell, external to the hydrogen burning shell. After the first dredge-up, at the luminosity bump, the external wing of the $^{7}$Be peak is the first to cross the molecular weight discontinuity. When it is connected with the convective envelope, $^{7}$Be produced via $^{3}He(\alpha,\gamma)^{7}Be$ starts to diffuse outwards. Depending on the mixing timescale, the Cameron \& Fowler (1971) mechanism may or may not operate in this region. As evolution proceeds, $^{7}$Be diffuses outwards and decays in the radiative layer where $^{7}$Li is destroyed by proton 
capture. In this thin region the energy released by $^{7}Li(p,\alpha)\alpha$ increases 
significantly. This could lead to a thermal instability strong enough to require a 
convective transport of the energy. As a consequence, $^{7}$Be could be instantaneously 
transported in the external convective envelope. This results in a sudden increase of 
the surface $^{7}$Li abundance. 
\newline
However, Palacios et al. have not yet published $^3$He yields and in the present work we adopted the $^{3}$He nucleosynthesis prescriptions from Sackmann \& Boothroyd (1999a,b).

\subsection{$^{4}$He}
All the stars eject part of their (old plus new) $^{4}$He content in the interstellar medium. The prescriptions about $^{4}$He nucleosynthesis are taken from van den Hoek \& Groenewegen (1997) for the low and intermediate  mass stars and from Thielemann, et al. (1996) for stars more massive than 10 M$_{\odot}$.

\subsection{The production matrix}
Recent chemical evolution models considering the evolution of the abundance of $^{3}$He are from 
Dearborn et al. (1996 - hereafter DST96), Prantzos (1996), Galli et al. (1997) 
and Palla et al. (2000). The models of Dearborn et al. (1996), 
tried to solve the problem posed by $^{3}$He overproduction in many ways. One way is allowing $^{3}$He to be reduced to a value of about $10^{-5}$ in lower mass stars, with cutoffs of 1$M_{\odot}$, or 1.6$M_{\odot}$. However, as pointed out by the authors, such low-mass stars have long lifetimes, and the reduction in $^{3}$He is only effective during recent epochs. Another way is to ignore the new $^{3}$He production by setting the surviving abundance factor $g3=X_{3_{f}}/X_{3_{i}}$ equal to 1 below 2 $M_{\odot}$, or between 1 and 2 $M_{\odot}$, where $X_{3_{i}}$ and $X_{3_{f}}$ are the initial and the final $^{3}$He stellar abundance, respectively. However, in this case is difficult to 
explain the planetary nebulae data (see Galli et al. 1997). 
\newline
Prantzos (1996) assumed the following prescriptions:
stars with $M>5M_{\odot}$ are always assumed to deplete their initial D + $^{3}$He, whereas lower mass stars are assumed to maintain their initial D plus $^{3}$He, plus the $^{3}$He produced by the $p~-~p$ chains. In this way, Prantzos (1996) succeeded in avoiding $^{3}$He overproduction just by setting to zero the $p~-~p$ contribution.
However, these prescriptions do not consider the dependence of extra-mixing on stellar mass and metallicity. 
\newline
In the present work we take into account extra-mixing, and its dependence from the stellar mass, metallicity as well as the $^{3}$He content of the proto-stellar cloud.
In order to account for the $^{3}$He nucleosynthesis prescription in our chemical evolution model, we adopt the formalism of Talbot \& Arnett (1971, 1973). In particular, we define the fraction of a star of mass $m$ which is ejected back into the interstellar medium in the form of isotopic species $i$, $R_{m_{i}}$,
\begin{eqnarray}
R_{m_{i}}(t)\equiv\sum_{i}Q_{m_{i,j}}X_{j}(t),
\end{eqnarray}
where $Q_{m_{i,j}}$ is a time independent ``production matrix'', and is defined for each stellar mass. The matrix specifies the mass fraction of the star in which material initially in the form of species $j$ is ejected as species $i$\footnote{Diagonal elements represent the mass fraction in which species $i$ is unprocessed and ejected.}, and $X_{i}(t)$ are the abundance mass fractions. $R_{m_{i}}(t)$ is a mass fraction, and should be positive, whereas the single elements $Q_{m_{i,j}}$ can also be negative. It is possible to think about negative $Q_{m_{i,j}}$ as the consequence of the fact that the species $i$ is mainly destroyed to form heavier species.
\newline
According to Talbot \& Arnett (1973), we define w2, which is the mass fraction ejected in the form of newly created D, and our choice is to set this quantity equal to zero, which means that all the D present in the stars is destroyed. The quantity q3 is instead the mass fraction within which any original $^{3}$He is converted to $^{4}$He or heavier species. Talbot \& Arnett (1973) definition of q3 is q3$=max(d,0.60)$ (based on the results of Iben 1967), where $d$ is the mass fraction which remains as a stellar remnant. Finally, the quantity w3 is the mass fraction ejected in the form of newly created $^{3}$He. 
\newline
Here we redefine q3 and w3 by including extra-mixing and hot-bottom burning in low and intermediate mass stars.

\subsubsection{Redefinition of the elements of the production matrix related to $^{3}$He}
$^{3}$He destruction factors, from Boothroyd (1999)\footnote{It has been used the CBP-lo case, see \texttt{http://www.krl.caltech.edu/$\sim$aib/} for more details.}, have been used to define q3 for stellar masses below $9 M_{\odot}$. 
\newline
The new definition of q3 is:
\begin{eqnarray}q3=1-g3,\end{eqnarray}
 where g3 = $X_{3_{f}}/X_{3_{i}}$. The initial abundance of $^{3}$He,
$X_{3_{i}}$, is a function of time and it depends on the chemical evolution model adopted. In fact, $X_{3_{i}}=X_{3_{i_{gas}}}+3/2\cdot~X_{2_{i_{gas}}}$, where $X_{3_{i_{gas}}}$ and $X_{2_{i_{gas}}}$ are respectively the interstellar gas abundances of $^{3}$He and D which are predicted by the model.
$X_{3_{f}}$ is the final stellar abundance of $^{3}$He, and it is influenced by possible extra-mixing and hot-bottom burning, depending on the stellar mass, metallicity and on the initial composition $X_{3_{i}}$. 
\newline
Attention should be payed to the dependence of g3 from the stellar abundance of $^{3}$He. The build-up of the $^{3}$He peak during the Main Sequence is independent from the initial $^{3}$He abundance, which is the $^{3}$He abundance in the protostellar nebula (following the pre main sequence burning). A fraction of the $^{3}$He produced during the MS phase is then added to the convective envelope after the dredge-up. However, if the initial $^{3}$He abundance was high, then the added $^{3}$He, from the pocket created during the MS phase, makes a smaller fractional increase relative to the initial value. 
Finally, at least in low mass stars, extra-mixing on the RGB destroys a fraction of the $^{3}$He present in the envelope at that time. In this case, the fraction destroyed depends mainly on the star mass and metallicity. Therefore:
\begin{eqnarray}^{3}He_{\rm final}=(^{3}He_{\rm initial}\cdot f_{\rm dredge-up} + ^{3}He_{MS~\rm pocket})\cdot f_{\rm extra-mixing}\end{eqnarray}
 where $f_{\rm dredge-up}$ is a depletion factor due to dredge-up on the RGB, and $f_{\rm extra-mixing}$ is a depletion factor due to extra-mixing (see Sackmann \& Boothroyd 1999a,b for further details). 
\newline
For  stellar masses larger than $9 M_{\odot}$ we adopted the Talbot \& Arnett (1973) q3 definition.
\newline
The quantity w3 has been also redefined, according to Sackmann \& Boothroyd (1999a). This quantity is a function of g3, which is defined between 0.85 and 9 M$_{\odot}$. For stellar masses larger than $9 M_{\odot}$, we define g3 = $1-q3$ (see the appendix for the detailed definition). For stars with masses below 2.5 M$_{\odot}$ we adopted
$w3=w3_{\rm CBP_{\rm Boothroyd}}$ for a given percentage $f$ of the stars in this mass range and $w3_{\rm standard_{\rm DST96}}$ for the remaining percentage of $(1-f)$ of the stars. In particular, the quantity $f$ is the percentage of stars in the low and intermediate mass range, which suffer from cool bottom processing (CBP), and $w3_{\rm standard_{\rm DST96}}$ is the quantity w3 predicted by standard stellar models. For masses larger than 2.5 M$_{\odot}$, $w3=w3_{CBP_{\rm Boothroyd}}$.
\newline
In Table 4 we report the various amounts of extra-mixing that we adopted in our models. Each prescription for extra-mixing is indicated by a letter (A, B, C, D). Model A is essentially the model presented in Chiappini \& Matteucci (2000), except for the fact that we are now adopting different yields for massive stars and a primordial helium abundance of 0.238 (Gruenwald et al. 2002) instead of 0.241 (Viegas et al. 2000). Model B assumes extra-mixing in 70\% of the stars with M$<$ 2.5 M$_{\odot}$, as suggested by Galli et al. (1997); Model C assumes extra-mixing in 93\% of stars with M$<$ 2.5 M$_{\odot}$, as suggested by Charbonnel \& Do Nascimento (1998) and their study of the carbon isotopic ratio $^{12}$C/$^{13}$C in field and cluster giants stars; Model D assumes extra-mixing in 99\% of the stars which is quite an extreme case. The case with 100\% extra-mixing is excluded by the observations of high $^{3}$He abundances in planetary nebulae such as NGC 3242 (Balser et al. 1999).
\begin{table}
\caption{Initial abundances of the models.}
\begin{tabular}{llllll}
Model & $10^{5}\cdot$(D/H)$_{\rm p}$ by number & $10^{5}\cdot $X$_{\rm D_{\rm p}}$ by mass & Y$_{\rm p}$ & $10^{5}\cdot(^{3}$He/H$)_{\rm p}$ by number & $10^{5}\cdot $X$_{^{3}\rm He_{\rm p}}$ by mass\\
\hline
&&&&&\\
I & 2.9 & 4.4 & 0.238 & 0.9 & 2\\
II & 2.5 & 3.8 & 0.248 & 1.0 & 2.3\\
III & 3.0 & 4.5 & 0.247 & 1.1 & 2.5\\
IV & 3.75 & 5.7 & 0.246 & 1.2 & 2.7\\
V & 10 & 15.2 & 0.240 & 2.0 & 4.6\\
VI & 30 & 46.3 & 0.229 & 3.0 & 6.9\\
&&&&&\\
\hline
\end{tabular}
\label{table3}
\end{table}
\begin{table}
\caption{Nucleosynthesis parameters of the models.}
\begin{tabular}{llll}
Model & q3 & w3 & extra-mixing\\
\hline
&&&\\
A & Talbot \& Arnett (1973) & DST96 & 93\%\\
B & new & new & 70\%\\
C & new & new & 93\%\\
D & new & new & 99\%\\                  
&&&\\
\hline
\end{tabular}
\label{table4}
\end{table}
\section{Results}
\subsection{The Solar Vicinity}
A good chemical evolution model should be in agreement with what is called the minimum set of observational constraints, among which the most important is the G-dwarf metallicity distribution (Chiappini et al. 1997; Kotoneva et al. 2002). In Table 5, the predicted and observed present-day quantities are shown. In particular the SN rate, the star formation rate, the surface gas density, the ratio between the surface gas density and the surface total density, the infall rate and the nova outburst rate. The H, D, $^{3}$He and $^{4}$He abundances at 9.5 Gyr (assuming 4.5 Gyr as the age of the Proto Solar Cloud) predicted by type I, II, III, IV models, are shown in Table 6. The depletion factors for D, and the enhancement factors for $^{3}$He and for $^{4}$He, $X_{t_{\rm Gal}}/X_{0}$, are shown in Table 7.  $X_{t_{\rm Gal}}$ is the abundance by mass of an element at the present time.

\begin{table}
\caption{Observed and predicted quantities at $R_{g,\odot}$ and $t = $14 Gyr. The predictions of the models do not differ since the difference concerns mainly the $^{3}$He production. It has been assumed a Galactic radius of 18 kpc, thus an area for the Galactic plane of $\approx 10^{9}$ pc$^{2}$.}
\begin{tabular}{lll}
  & Models & Observed (see Matteucci 2001)\\
\hline
&&\\
SNIa (century$^{-1}$) & 0.3 & 0.6h$^{2}$\\
SNII (century$^{-1}$) & 0.9 & 0.8h$^{2}$\\
$\Psi$(R$_{g,\odot}$,t$_{now}$) (M$_{\odot}$ pc$^{-2}$ Gyr$^{-1}$) & 2.6 & 2 - 10\\
$\sigma_g$(R$_{g,\odot}$,t$_{now}$) (M$_{\odot}$ pc$^{-2}$) & 7 & 6.6 $\pm 2.5$\\
$\sigma_g$ / $\sigma_T$ (R$_{g,\odot}$,t$_{now}$) & 0.13 & 0.09 - 0.21\\
$\dot{\sigma}_{inf}$(R$_{g,\odot}$,t$_{now}$)(M$_{\odot}$ pc$^{-2}$ Gyr$^{-1}$) & 1 & 0.3 - 1.5\\
Novae Outbursts (yr$^{-1}$) & 25 & 20 - 30\\
&&\\
\hline\\
\end{tabular}
\label{Table5}
\end{table}
\begin{table} 
\caption{Abundances at 9.5 Gyr, by mass.The observations are from (a) Anders \& Grevesse (1989)
and (b) Geiss \& Gloeckler (1998).}
\begin{tabular}{llllllll}
Element & A-I & C-I & C-II & C-III & C-IV & D-I & Observations\\
\hline
&&&&&&&\\
H & 0.73 & 0.73 & 0.72 & 0.72 & 0.72 & 0.73 & 0.71 $\pm$ 0.02 (a)\\
D$\cdot 10^{5}$ & 3.4 & 3.4 & 2.9 & 3.5 & 4.3 & 3.4 & 3.0 $\pm$ 0.5 (b)\\
$^{3}$He$\cdot 10^{5}$ & 2.2 & 3.0 & 3.1 & 3.3 & 3.6 & 2.7 & 3.2 $\pm$ 0.2 (b)\\
$^{4}$He & 0.252 & 0.252 & 0.262 & 0.261 & 0.260 & 0.252 & 0.274 $\pm$ 0.016 (a)\\
&&&&&&&\\
\hline
\end{tabular}
\label{Table6}
\end{table}
\begin{table} 
\caption{Depletion factors, for D ($X_{0}/X_{t_{\rm Gal}}$), and enhancement factors for $^{3}$He and $^{4}$He ($X_{t_{\rm Gal}}/X_{0}$) at the solar vicinity.}
\begin{tabular}{llllllll}
Element & A-I & C-I & C-II & C-III & C-IV & D-I\\
\hline
&&&&&&\\
D & 1.5 & 1.5 & 1.5 & 1.5 & 1.5 & 1.5\\
$^{3}$He & 2.5 & 3.0 & 2.7 & 2.6 & 2.4 & 1.7\\  
$^{4}$He & 1.1 & 1.1 & 1.1 & 1.1 & 1.1 & 1.1\\                                           
&&&&&&&\\
\hline
\end{tabular}
\label{Table7}
\end{table}
\subsubsection{Abundances of Deuterium}
In Fig. 1 (upper panel) we show the evolution of D/H with time, as predicted by our models I, II, III and IV.
The predictions seem to fit the data well for the PSC and the LISM, although the data for the LISM show a large spread. The data are from Linsky (1998), from the seven lines of sight observed from the FUSE satellite (Moos et al. 2000, 2002) and from other lines of sight observed by Copernicus, IMAPS, IUE and HST (see Table 4 in Moos et al. 2002 and references therein). There is evidence of spatial variability for the D abundances in the ISM beyond 0.1 kpc (see the data from FUSE observations of Feige 110, Friedman et al. 2002, (D/H)$=2.14\pm 0.82\cdot 10^{-5}$, IMAPS observations of $\delta$ Orionis, Jenkins et al. 1999, (D/H)$=0.74^{+0.12}_{-0.09}\cdot 10^{-5}$\footnote{Similar results are found by Bertoldi et al. (1999), who detect an infrared transition within the electronic ground state of the deuterated hydrogen molecule, HD.}, and $\gamma^{2}$ Velorum, Sonneborn et al. 2000, (D/H)$=2.18^{+0.22}_{-0.19}\cdot 10^{-5}$). 
\newline
Strangely enough, the underabundance of D toward $\delta$ Ori A is not accompanied by an overabundance of N or O relative to H, as expected because of the fact that destruction of D implies creation of N and O (see Steigman 2002). 
\newline
The reason of this dispersion in the data is not yet clear. Perhaps some elements are systematically removed from the gas phase as they are incorporated into interstellar dust (Savage \& Sembach 1996), but this does not seem the case for N (Meyer et al. 1997). 
Another possibility is to call for infalling material, as in the case of this model, and for various efficiencies of mixing processes within the ISM (see Moos et al. 2002 and Lemoine et al. 1999 for a discussion of the possible reasons for the interstellar D/H variations).
\newline
In Fig. 1 (lower panel), we show the predictions of models V and VI with high primordial D abundance. It is clear from the figure that a high primordial D abundance should be ruled out since it predicts a too high D abundance for the PSC and the LISM, even taking into account possible spatial variability for the LISM D abundance.
Therefore, cases V and VI are ruled out (see also Tosi et al. 1998). 
\newline
Fig. 2 shows the evolution of D abundance as a function of [Fe/H] at different galactocentric distances, as predicted by model C-I. It is clear from the figure that the model depletes much more D near the Bulge, whereas D should be quite close to the primordial value in the outer regions of the Galaxy (see Sect. 4.2). Chengalur et al. (1997) detected a low significance feature of DI implying (D/H)$=3.9 \pm 1.0\cdot 10^{-5}$ at large galactocentric distances. This value is close to the primordial value and consistent with the low D/H measured towards some of the high redshift absorbers. Three recent measurements of D towards the galactic center are from Polehampton et al. (2002), Lubowich et al. (2000) and Jacq et al. (1999). 
Polehampton et al. (2002) observed the giant molecular cloud Sagittarius B2,
located near the Galactic Centre, in the far-infrared. They obtained 
(D/H) in the range (0.2 $-$ 11) $\cdot 10^{-6}$.
Lubowich et al. (2000) reported the detection of D in a molecular cloud at only 10 pc from the Galactic center. They estimated a (D/H)$=1.7 \pm 0.3\cdot 10^{-6}$, around 10 times below the local interstellar value. Essentially the same
value is given by Jacq et al. (1999). Apart from these measurements in the Galactic bulge, 
there are no more measurements in the inner regions of the Galactic disk.

\subsubsection{Abundances of $^{3}$He}
Figures 3 and 4 show the time-evolution of $^{3}$He. In Fig. 3 we show models A, B, C and D computed with initial abundances of type I (see Table 3). Type A models, with 93\% extra-mixing, old q3 and w3 definition do not fit the PSC data very well (Geiss \& Gloeckler 1998). Type B models, with 70\% extra-mixing, new q3 and w3 definition, clearly overproduces $^{3}$He with respect to both PSC and ISM observed abundances. Type C models, with 93\% extra-mixing, new q3 and w3 definition, show an earlier rise and are in better agreement with the PSC data, whereas type D models, with 99\% extra-mixing, new q3 and w3 definition, are quite flat. Fig. 4 shows how the time-evolution depends on the initial abundance when the nucleosynthesis prescriptions are fixed. Case V, with (D/H)$_{\rm p} = 10\cdot 10^{-5}$ which implies ($^{3}$He/H)$_{\rm p} = 2.0\cdot 10^{-5}$, disagrees not only with the PSC and LISM D abundances, but also with those of $^{3}$He. The same is true for case VI. All models, except type B models, seem to be acceptable, thus indicating that a percentage of extra-mixing larger than 70 \% should be assumed (see also Galli et al. 1997).

\subsubsection{Abundances of $^{4}$He}
The predicted solar $^4$He abundance by type II, III and IV models are in good agreement 
with the observations. Type I models instead, predict a slightly lower $^4$He abundance.

In Fig. 5 we show the model predictions for Y versus O/H. 
We obtain, for model C-IV, $\Delta Y/ \Delta (O/H)$ = 52 for 360 $\leqslant 10^6$(O/H)$ \leqslant$ 660. This value is in good agreement with the value of 50.2 $\pm$ 3.9 recently reported by Maciel (2001), based on a planetary nebula sample of 81 objects. We notice that, although planetary nebulae are interesting objects to study the helium evolution in the Galaxy, these objects can have contaminated helium abundances. Maciel (2001) took into account the He contamination by planetary nebulae progenitor stars and obtained a lower value than the previous $\Delta Y/ \Delta (O/H)$ and $\Delta Y/ \Delta Z$ estimates based on the same kind of objects (e.g. Chiappini \& Maciel 1994).

Our C-IV model predicts $\Delta Y/ \Delta Z$ $\simeq$ 1.2 if the entire metallicity range is considered, whereas it is $\simeq$ 2.75 for $0.016\lesssim Z\lesssim 0.018$. For $0.008\lesssim Z\lesssim 0.018$, $\Delta Y/ \Delta Z\simeq 1.5$. The $\Delta Y/ \Delta Z$ ratio is usually determined from photoionized nebulae such as HII regions and HII galaxies (e.g. Lequeux et al. 1979; Pagel et al. 1992; Pagel 2000). More recent results are from Esteban et al. (1999) who obtained $\Delta Y/ \Delta Z$ = 2.87 $\pm$ 1.0, 2.81 $\pm$ 1.0 and 1.87 $\pm$ 0.5 for Orion, M8 and M17 respectively. These authors adopted Y$_{\rm p} = $ 0.240 $\pm$ 0.005. This estimate is consistent with that of Pagel \& Portinari (1998), who found $\Delta Y/ \Delta Z$ = 3$\pm$2, from the analysis of the metallicity dependent location of the main sequence of nearby stars from Hipparcos data for stars with Z $\leqslant Z_{\odot}$. 
Finally, Maciel (2001) estimates 2.0 $\leqslant \Delta Y/ \Delta Z \leqslant$ 2.7 for Y$_{\rm p} = $ 0.24 and 2.9 $\leqslant \Delta Y/ \Delta Z \leqslant$ 3.6 for Y$_{\rm p} =$ 0.23. The predictions of model C-IV are in good agreement with the above observations.

\subsection{The Galactic Disk} 
In this section we show our results for the predicted radial abundance distributions of D, $^{3}$He and $^{4}$He. In Figures 6 and 7 we show the predicted gradients of D/H, D/O and D/N (the local values of these ratios in the ISM are also shown).
The predicted gradient for D/H is positive and steep, due to the fact that D is destroyed in stars (see also Prantzos 1996). The evolution of the inner disk regions are faster than the outer parts, which are still in the process of forming thus having an almost primordial composition (Fig. 6). 
In Fig. 6 (upper panel) we show the three measurements of D in the Galactic center which show a factor of $\simeq$ 10 depletion with respect to the solar vicinity value. We predict a depletion factor of $\simeq$ 5.7 at 2 kpc from the Galactic center and $\simeq$ 2 at a distance of 4 kpc. For large galactocentric distances ($>$ 14 kpc) a value close to the primordial one is expected.
In fact, D is probably the most sensitive element to radial variations in the timescale of disk formation. The gradients for D/O and D/N are even steeper, because when D is destroyed by stellar nucleosynthesis, O and N are produced as a consequence (Fig. 7). 
The predicted gradients for D/H, D/O and D/N are $d\log$(D/H)$/d$R $\simeq$ 0.02 dex/kpc, $d\log$(D/O)$/d$R $\simeq$ 0.13 dex/kpc, $d\log$(D/N)$/d$R $\simeq$ 0.14 dex/kpc, respectively, over the 4-18 kpc range. 
\newline
In Figures 8 and 9 the predicted $^{3}$He abundance distribution along the disk is shown. For this element, we predict a negative gradient, since it is more produced than destroyed. In the inner regions, which are older in the inside-out scenario, there has been more stellar activity, and the contribution of low and intermediate mass stars to the $^{3}$He enrichment of the ISM has been more important than in the outer regions. 
Fig. 8 shows how the $^{3}$He distribution is dependent from the percentage of extra-mixing, whereas Fig. 9 shows how the distribution is dependent from the adopted ``primordial'' $^{3}$He abundance, once the extra-mixing percentage is fixed. It can be seen that the assumption that 99\% of the low and intermediate mass stars suffer extra-mixing leads to a flatter distribution, with $-0.02<d\log$($^{3}$He/H)$/d$R$<-0.01$ dex/kpc. The case with 93\% extra-mixing predicts $-0.04<d\log$($^{3}$He/H)$/d$R$<-0.03$ dex/kpc, in agreement with the data from Galactic HII regions (Bania et al. 2002). 
\newline
Finally, in Fig. 10 the $^{4}$He distribution along the disk is shown, with a predicted small gradient $d$($^{4}$He/H)$/d$R $\approx -$0.002 kpc$^{-1}$ over the 4-18 kpc range. The shaded region shows the locus of the planetary nebulae data (see Maciel 2001 for the details). 
Planetary nebulae data extend up to $\simeq$ 12 kpc and show an essentially flat distribution along the disk.
\newline
Taking all of these results into account, we conclude that our best model is C-IV, which assumes an extra-mixing percentage of 93\%, which is also
 the percentage suggested by observations (Charbonnel \& do Nascimento 1998).
This model also assumes the primordial D/H abundance of $3.75\cdot 10^{-5}$, which implies ($^{3}$He/H)$_{\rm p}=1.2\cdot 10^{-5}$ and $Y_{\rm p}=0.246$. The primordial abundance of deuterium is from Levshakov et al. (2002). These authors measured one of the highest value for the deuterium abundance within the set of ``low-D/H'' observations of high redshift absorbers.
 
\section{Conclusions}
In this paper we have calculated the galactic evolution of the abundances of D, $^{3}$He, and $^{4}$He. In particular, we modelled the evolution of the $^{3}$He abundance taking into account in detail the extra-mixing process in low and intermediate mass stars, and its dependence on stellar mass and metallicity. We have predicted the time-evolution of the abundances of D, $^{3}$He, and $^{4}$He, and their distribution along the Galactic disk, and studied how these abundances depend upon the ``primordial'' abundances as well as upon the percentage of extra-mixing adopted in the models. We have compared these predictions with the most recent observations, including the first published data from the FUSE mission and selected the best combination of parameters.

The adopted chemical evolution model, the ``two-infall'' one, differs from other models present in the literature because it assumes that the galactic thin-disk formed out of an infall episode completely disentangled from the formation of the galactic halo, with almost no contribution from the pre-enriched halo gas especially in the inner regions. As a consequence of this, our predicted deuterium depletion factors are lower than in previous models (e.g. Prantzos 1996), especially in the region close to the bulge. The reason for this resides in the fact that after the first halo phase, where some D depletion already occurs, the formation of the thin-disk implies mostly primordial gas and therefore the D abundance in the gas increases again and then declines (see Tosi et al. 1998).
For what concerns the predicted abundances of $^{3}$He our model
differs from previous ones because it includes new nucleosynthesis prescription
for this element.
\newline
The main conclusions are the following:
\begin{itemize}
\item[-] The redefinition of the production/destruction of $^{3}$He in low and intermediate mass stars has important consequences on the evolution of the $^{3}$He abundance and on its distribution along the Galactic disk. In particular, the new models with 93\% of extra-mixing, show a good agreement with the PSC observations as well as with the observed gradient along the disk. We predict a $^{3}$He gradient of $-0.04<d\log$($^{3}$He/H)/$d$R$<-0.03$ dex/kpc in the galactocentric distance range 4-18 kpc.

\item[-] The predicted time evolution of D in the solar vicinity is well reproduced although the LISM values measured by FUSE show a large spread. Since D is only destroyed during galactic evolution, we predict a positive gradient of this element along the disk, in particular $d\log$(D/H)$/d$R$\simeq$ 0.02 dex/kpc, $d\log$(D/O)$/d$R$\simeq$ 0.13 dex/kpc, $d\log$(D/N)$/d$R$\simeq$ 0.14 dex/kpc, in the galactocentric distance range 4-18 kpc. Our predicted D/H gradient is less steep than the one obtained by Prantzos (1996) by means of a model with a much lower infall rate than the ``two-infall'' model. This is a consequence of the fact that D abundance is very sensitive to the infall/star formation history in the Galaxy. It should be noted that the infall/star formation history assumed here is requested to fit the majority of the observational constraints (Chiappini et al. 1997, 2001). We predict that observations towards the Galactic anticenter should provide a D abundance close to the primordial value.
We predict a depletion factor for D in the solar vicinity of $\simeq$ 1.5, whereas at 2 kpc from the Galactic center we expect a factor of $\simeq$ 5. However, the model of the Galactic disk is probably not appropriate to describe the Galactic bulge which should have suffered a much higher star formation than we can deduce from the present models (see Matteucci et al. 1999).

\item[-] Chemical evolution models can constrain the primordial value of the deuterium abundance and a value of (D/H)$_{\rm p} \lesssim$ 4 $\cdot$ 10$^{-5}$ is suggested by the present model. This upper limit implies, according to Standard Big Bang Nucleosynthesis (Burles et al. 1999), $\Omega_{b}h^2 \gtrsim 0.017$, in agreement with estimates from the Cosmic Microwave Background radiation analysis, which implies $\Omega_{b}h^{2}_{\rm CMB} = 0.032^{+0.009}_{-0.008}$, at 95\% Confidence Level (Jaffe et al. 2001). This limit seems to rule out high D abundance detection in DLA absorbers, and implies $Y_{\rm p} \geq 0.244$ from SBBN. This is at variance with the latest estimate of primordial $^{4}$He from Gruenwald et al. (2002), who suggest that the primordial $^{4}$He should be Y$_{\rm p}$ = 0.238$\pm$0.003 (see also  Sauer \& Jedamzik 2002). Finally, we predict an enrichment of $^{4}$He relative to metallicity of $\Delta Y/ \Delta Z=1.5$ in the range $0.008\leqslant Z\leqslant 0.018$, and a gradient along the galactic disk $d\log$($^{4}$He/H)$/d$R $\approx -0.007$ dex/kpc, or $d$($^{4}$He/H)$/d$R $\approx -0.002$ kpc$^{-1}$, in the galactocentric distance range 4-18 kpc.
\end{itemize}

\begin{acknowledgements}
We thank A. Boothroyd for clarifying us some details of his stellar model
predictions. We also thank the referee, D. Balser, and G. Steigman for the many suggestions
which improved this work.
\end{acknowledgements}

\newpage

{\appendix\section{detailed definition of the quantity w3.}
The quantity w3, the mass fraction ejected in the form of newly created $^{3}$He, as defined from DST96 is: 
\newline
if 0.65 $M_{\odot}<M_{*}<2.5$ $M_{\odot}$,
\begin{eqnarray}
k_{a}&=&(0.00007+0.00135\cdot M_{*}^{-2.2})\nonumber\\
k_{b}&=&(0.55-0.3\cdot log_{10} M_{*})\nonumber\\
X_{3,*}&=&X_{3,i}+\frac{3X_{2,i}}{2}\nonumber\\
w3&=&\frac{k_{a}+k_{b}(X_{3,*}-2.1\cdot 10^{-4})}{M_{*}\cdot 0.77};\nonumber
\end{eqnarray}
if 2.5 $M_{\odot}<M_{*}<25$ $M_{\odot}$,
\begin{eqnarray}
k_{a}&=&(0.000111+\frac{0.00085}{M_{*}^{2}}-0.5\cdot 10^{-6}\cdot M_{*}^{1.5})\nonumber\\
k_{b}&=&(0.485-0.022\cdot e^{\frac{M_{*}}{10}})\nonumber\\
k_{c}&=&1.7\cdot 10^{-5} e^{-(\frac{X_{3,*}}{7.5\cdot 10^{-6}})^{2}}\nonumber\\
w3&=&\frac{k_{a}+k_{b}(X_{3,*}-2.1\cdot 10^{-4})-k_{c}}{M_{*}\cdot 0.77};\nonumber
\end{eqnarray}
if 25 $M_{\odot}<M_{*}$,
\begin{eqnarray}
k_{a}&=&(0.00011-4\cdot 10^{-7}M_{*})\nonumber\\
k_{b}&=&0.332\nonumber\\
w3&=&\frac{k_{a}+k_{b}(X_{3,*}-2.1\cdot 10^{-4})}{M_{*}\cdot 0.77},\nonumber
\end{eqnarray}
where $M_{*}$ is the stellar mass, $X_{3,*}$ is the initial stellar $^{3}$He abundance by mass (after the pre-main-sequence burning of D into $^{3}$He - see for example Mazzitelli \& Moretti 1980). 
\newline
The quantity w3 as defined by Boothroyd (1999), over the entire mass range, is a function of g3, which is defined as $g3=X_{3_{f}}/X_{3_{i}}$ between 0.85 and 9 M$_{\odot}$, and is a function of $X_{3,*}$, $M_{*}$, and the stellar metallicity Z. 
\newline
If 0.85 $M_{\odot}<M_{*}<4$ $M_{\odot}$,
\begin{eqnarray}
w3&=&\frac{X_{3,*}\cdot g3(X_{3,*},~M_{*},~Z)\cdot (0.95 M_{*}-0.5)}{M_{*}};\nonumber
\end{eqnarray}
if 4 $M_{\odot}<M_{*}<12$ $M_{\odot}$,
\begin{eqnarray}
w3&=&\frac{X_{3,*}\cdot g3(X_{3,*},~M_{*},~Z)\cdot (0.9 M_{*}-0.3)}{M_{*}};\nonumber
\end{eqnarray}
if 12 $M_{\odot}<M_{*}<12$ $M_{\odot}$,
\begin{eqnarray}
w3&=&\frac{X_{3,*}\cdot g3(X_{3,*},~M_{*},~Z)\cdot (0.982 M_{*}-1.22)}{M_{*}}.\nonumber
\end{eqnarray}
For stellar masses greater than $9 M_{\odot}$, we define g3 as $1-q3$, where q3 is the mass fraction within which any original\footnote{That is, within the protostellar nebula.} $^{3}$He is converted to $^{4}$He or heavier species.}

\newpage

\begin{figure} 
\resizebox{\hsize}{!}{\includegraphics{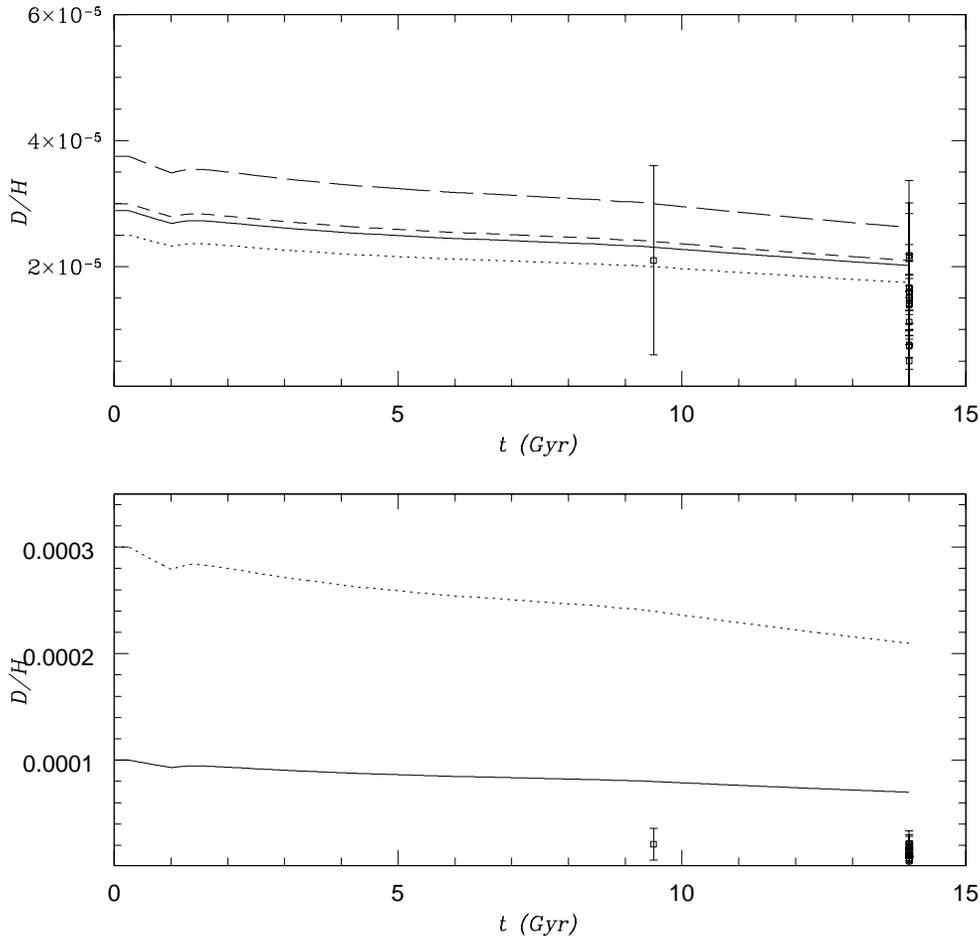}}
\caption{Predicted and observed (at 3$\sigma$) abundances of D as function of time (Gyr): Proto Solar Cloud data are from Geiss \& Gloeckler (1998), and LISM data are from FUSE mission observations along seven lines of sight (see Moos et al. 2002); observations from other lines of sight (see table 4 from Moos et al. 2002 and references therein) are also shown. The LISM observations are quite sparsed and there is evidence for spatial variability of the D abundances (see Jenkins et al.  1999; Sonneborn et al. 2000; Friedman et al. 2002).
\newline 
Model predictions are labeled as follows:
upper panel -- C-I (solid), (D/H)$_{\rm p} = 2.9\cdot 10^{-5}$; C-II (dotted), (D/H)$_{\rm p} = 2.5\cdot 10^{-5}$; C-III (short-dashed), (D/H)$_{\rm p} = 3.0\cdot 10^{-5}$; C-IV (long-dashed), (D/H)$_{\rm p} = 3.75\cdot 10^{-5}$;
lower panel -- C-V (solid), (D/H)$_{\rm p} = 10\cdot 10^{-5}$; C-VI (dotted), (D/H)$_{\rm p} = 30\cdot 10^{-5}$. 
}
\label{Fig1}
\end{figure}

\begin{figure}
\resizebox{\hsize}{!}{\includegraphics{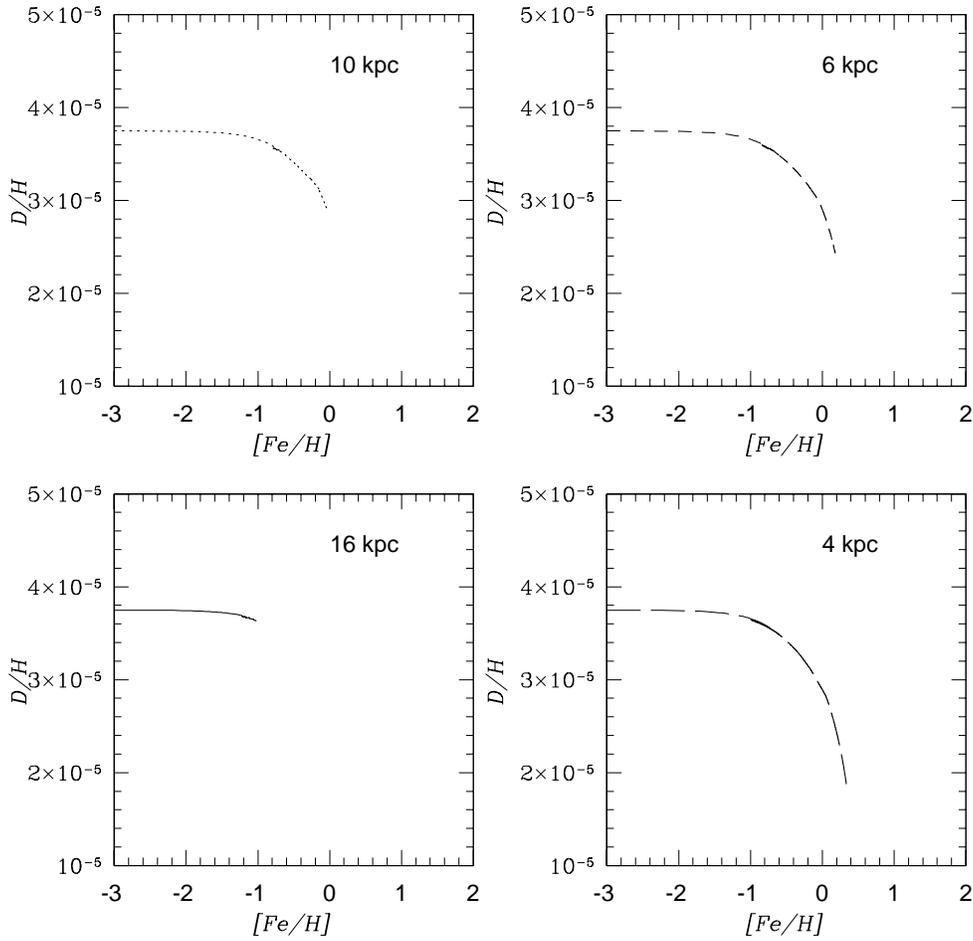}}
\caption{Abundances of D as a function of [Fe/H], at different galactocentric distances: R = 4 kpc (long-dashed), 6 kpc (short-dashed), 10 kpc (dotted), 16 kpc (solid). We show the predictions of model C-IV which assumes 93\% extra-mixing, new q3 and w3 definition and (D/H)$_{\rm p} = 3.75\cdot 10^{-5}$. There is more D depletion near the Bulge. At 16 kpc there is almost no depletion and the predicted value at 14 Gyr is close to the primordial value. The predictions from models A, B and D do not differ from the one shown here since the difference concerns mainly the $^{3}$He nucleosynthesis.}
\label{Fig2}
\end{figure}

\begin{figure}
\resizebox{\hsize}{!}{\includegraphics{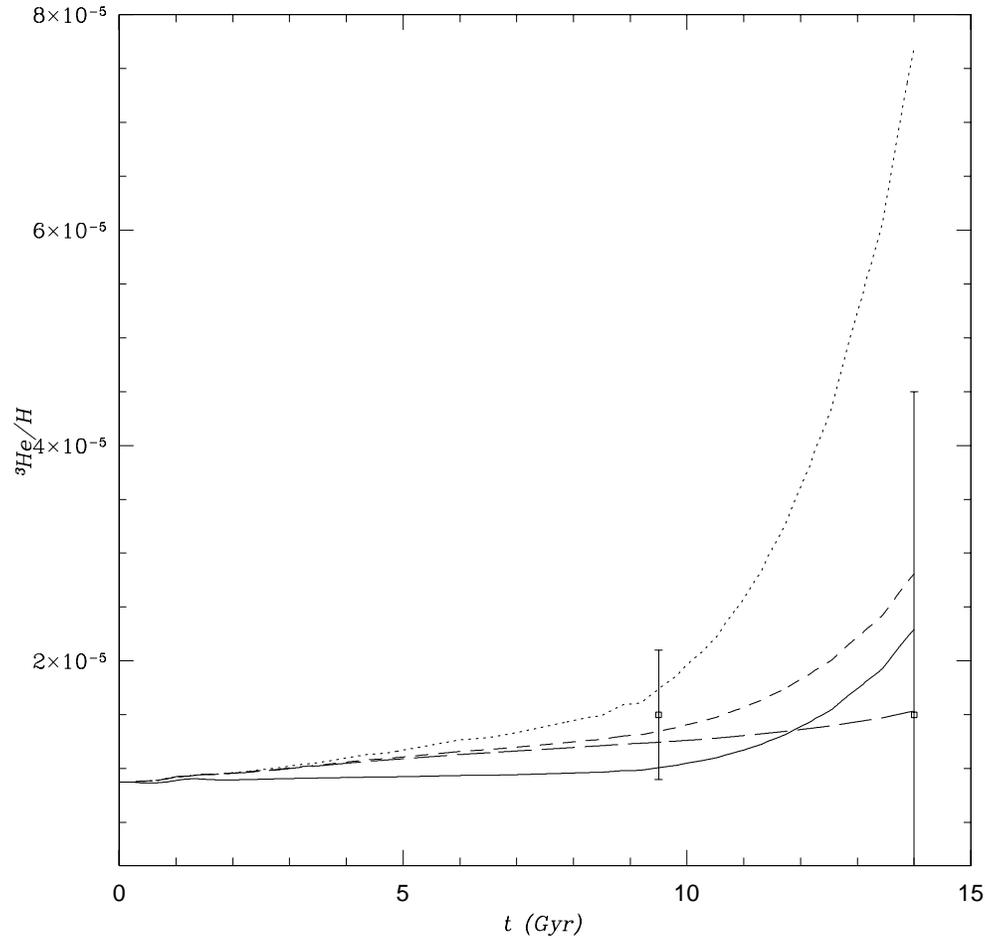}}
\caption{Abundances of $^{3}$He as a function of time (Gyr) as predicted by models assuming different amounts of extra-mixing: model A-I (solid), with 93\% extra-mixing, old q3 and w3 definition; model B-I (dotted), with 70\% extra-mixing, new q3 and w3 definition; model C-I (short-dashed), with 93\% extra-mixing, new q3 and w3 definition; model D-I (long-dashed), with 99\% extra-mixing, new q3 and w3 definition.
\newline
Also shown are the Proto Solar Cloud and ISM measured abundances (at 3$\sigma$) from Geiss \& Gloeckler (1998) and Linsky (1998). The enhancement factors are $\simeq 2.5$ for the model A-I, $\simeq 3.0$ for the model C-I and $\simeq 1.7$ for the model D-I. Model B-I clearly overproduces $^{3}$He. Models B, C and D, with redefined q3 and w3, show an earlier rise compared to model A.}
\label{Fig3}
\end{figure}

\begin{figure}
\resizebox{\hsize}{!}{\includegraphics{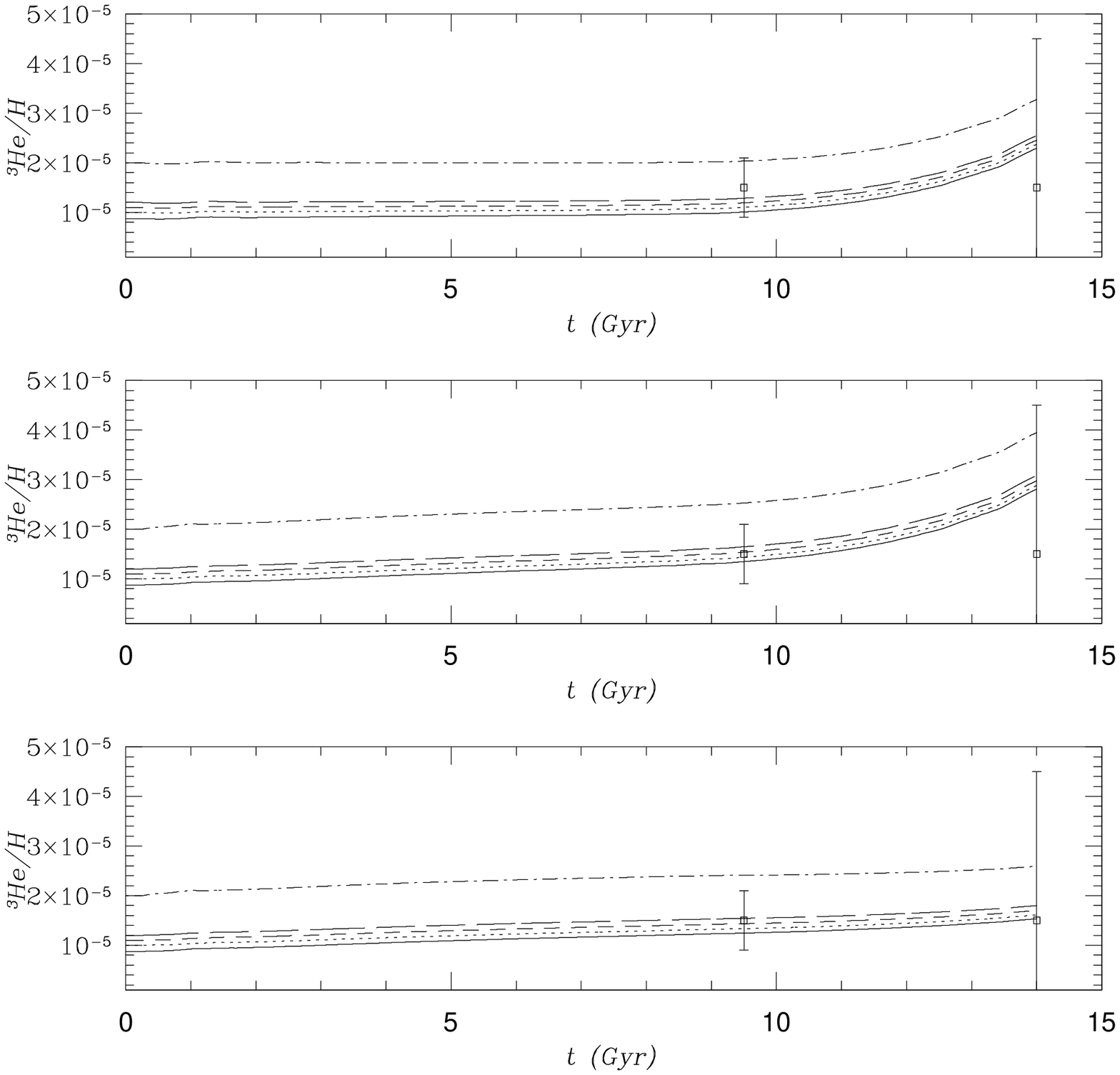}}
\caption{Abundances of $^{3}$He as a function of time (Gyr) as predicted by
models A, C and D. 
\newline
Lower panel: predictions of type-D models, with 99\% extra-mixing and new q3 and w3 definition. The curves are labelled as: D-I (solid), ($^{3}$He/H)$_{\rm p} = 0.9\cdot 10^{-5}$; D-II (dotted), ($^{3}$He/H)$_{\rm p} = 1.0\cdot 10^{-5}$; D-III (short-dashed), ($^{3}$He/H)$_{\rm p} = 1.1\cdot 10^{-5}$; D-IV (long-dashed), ($^{3}$He/H)$_{\rm p} = 1.2\cdot 10^{-5}$; D-V (dot-short-dashed), ($^{3}$He/H)$_{\rm p} = 2.0\cdot 10^{-5}$. 
\newline
Middle panel: predictions of type-C models, with 93\% extra-mixing and new q3 and w3 definition. The curves are labelled as: C-I (solid), ($^{3}$He/H)$_{\rm p} = 0.9\cdot 10^{-5}$; C-II (dotted), ($^{3}$He/H)$_{\rm p} = 1.0\cdot 10^{-5}$; C-III (short-dashed), ($^{3}$He/H)$_{\rm p} = 1.1\cdot 10^{-5}$; C-IV (long-dashed), ($^{3}$He/H)$_{\rm p} = 1.2\cdot 10^{-5}$; C-V (dot-short-dashed), ($^{3}$He/H)$_{\rm p} = 2.0\cdot 10^{-5}$. 
\newline
Upper panel: predictions of type-A models, with 93\% extra-mixing and old q3 and w3 definition. The curves are labelled as: A-I (solid), ($^{3}$He/H)$_{\rm p} = 0.9\cdot 10^{-5}$; A-II (dotted), ($^{3}$He/H)$_{\rm p} = 1.0\cdot 10^{-5}$; A-III (short-dashed), ($^{3}$He/H)$_{\rm p} = 1.1\cdot 10^{-5}$; A-IV (long-dashed), ($^{3}$He/H)$_{\rm p} = 1.2\cdot 10^{-5}$; A-V (dot-short-dashed), ($^{3}$He/H)$_{\rm p} = 2.0\cdot 10^{-5}$. 
\newline
These plots show the dependence of the time-evolution of the abundance of $^{3}$He/H from the initial abundance, once the nucleosynthesis prescriptions are fixed.} 
\label{Fig4}
\end{figure}

\begin{figure}
\resizebox{\hsize}{!}{\includegraphics{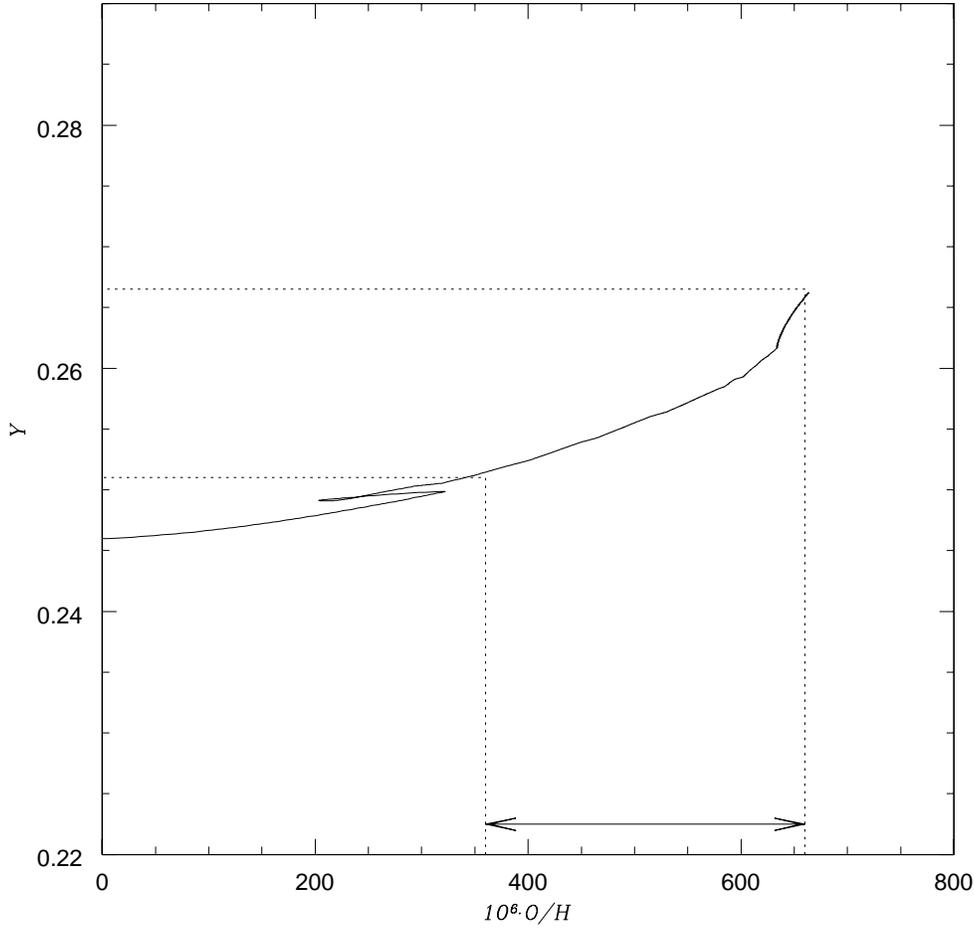}}
\caption{Model C-IV prediction for Y vs 10$^6$ O/H. The predicted value for $\Delta Y/ \Delta$(O/H) in the indicated range is $\simeq$ 52. Maciel (2001) estimated a $\Delta Y/ \Delta$(O/H) = 50.2 $\pm$ 3.9 for Y$_{\rm p} =$ 0.24 based on a sample of 81 planetary nebulae.}
\label{Fig5}
\end{figure}

\begin{figure}
\resizebox{\hsize}{!}{\includegraphics{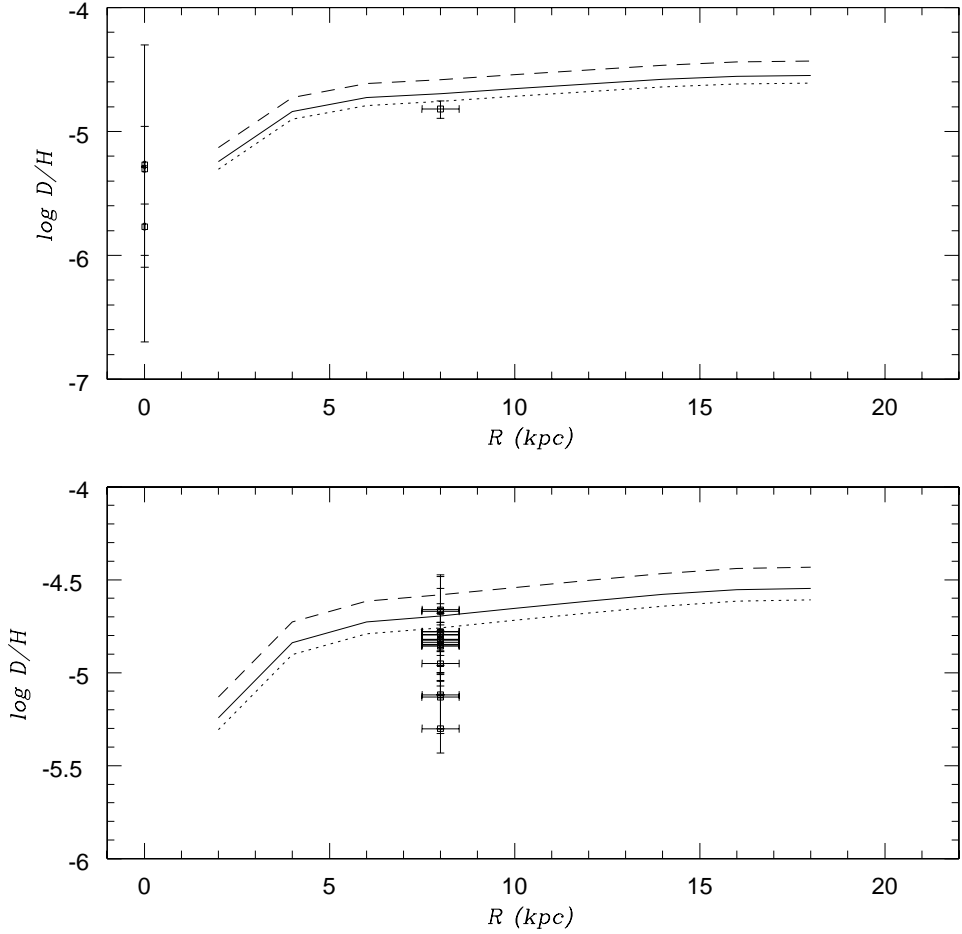}}
\caption{Distribution along the disk of D as a function of the Galactocentric distance, at t = 14 Gyr. Models C, with 93\% extra-mixing, new q3 and w3 definition are shown. Curves are labelled as follows: model C-I (solid), (D/H)$_{\rm p} = 2.9\cdot 10^{-5}$; C-II (dotted), (D/H)$_{\rm p} = 2.5\cdot 10^{-5}$; C-IV (short-dashed), (D/H)$_{\rm p} = 3.75\cdot 10^{-5}$.
\newline
Upper panel: the data point at R = 8 kpc is the weighted mean from FUSE mission observations (Moos et al. 2002 -- at $3\sigma$), along seven lines of sight. The three data points in the Galactic center are from Polehampton et al. (2002), Lubowich et al. (2000) and Jacq et al. (1999).
\newline 
Lower panel: the seven lines of sight from FUSE mission observations, Moos et al. (2002) and data, at $3\sigma$, from other lines of sight observed by Copernicus, IMAPS, IUE and HST are shown. The predicted gradient for all models is $d\log$(D/H)$/d$R$\approx 0.02$ dex/kpc over the 4-18 kpc range. }
\label{Fig6}
\end{figure}

\begin{figure}
\resizebox{\hsize}{!}{\includegraphics{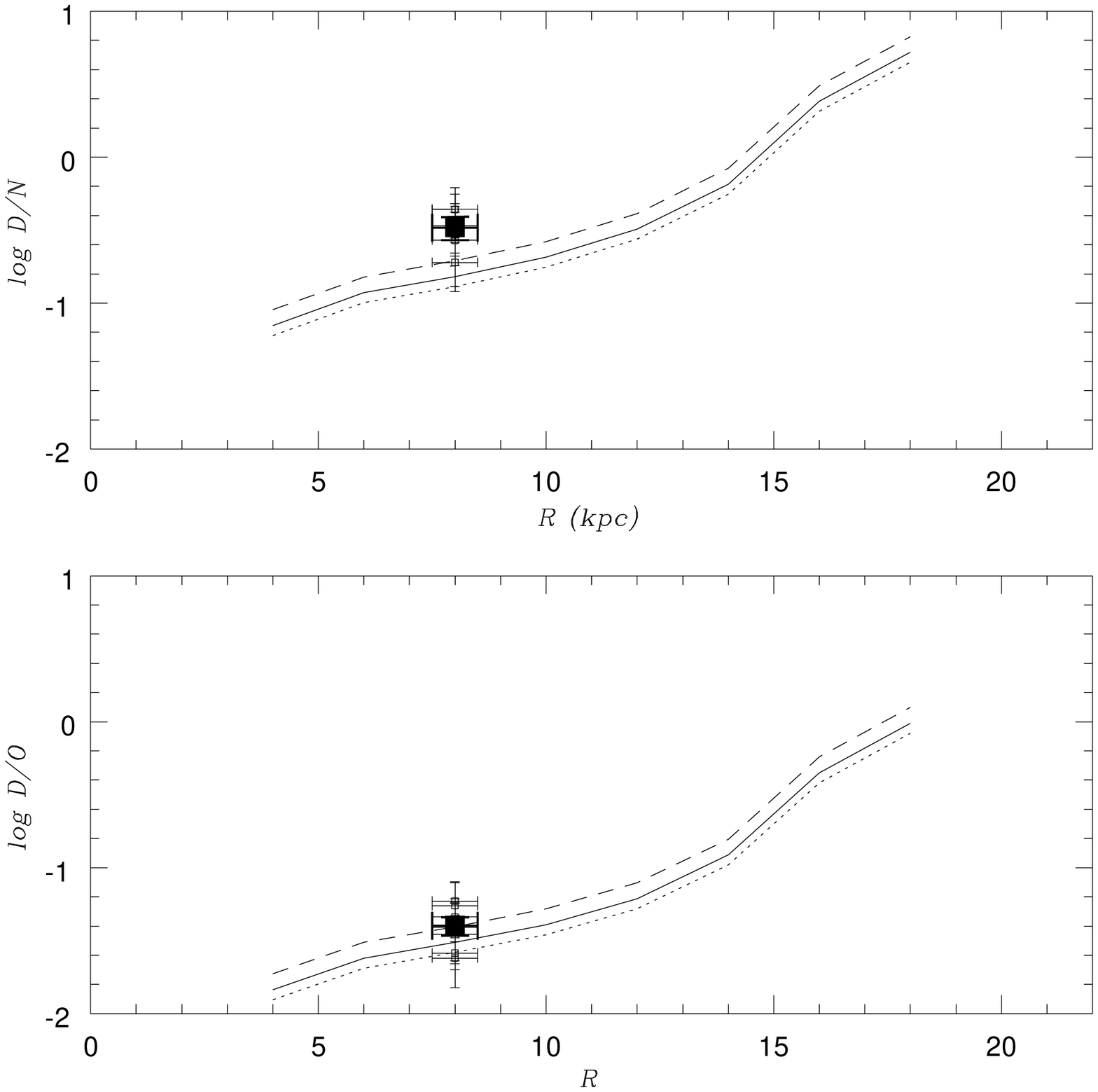}}
\caption{Distribution along the disk of $D/^{16}O$ and $D/^{14}N$, at t = 14 Gyr. Models are labelled as follows: model C-I (solid), (D/H)$_{\rm p} = 2.9\cdot 10^{-5}$; C-II (dotted), (D/H)$_{\rm p} = 2.5\cdot 10^{-5}$; C-IV (short-dashed), (D/H)$_{\rm p} = 3.75\cdot 10^{-5}$. 
\newline
The large symbol represents the weighted mean from FUSE mission observations (Moos et al. 2002 -- at $3\sigma$) which probe the LISM within 0.5 kpc. 
The predicted gradient for all models is $d\log$(D/O)$/d$R$\approx$ 0.13 dex/kpc and
$d\log$(D/N)$/d$R$\approx 0.14 dex/kpc$
over the 4-18 kpc range.}
\label{Fig7}
\end{figure}

\begin{figure}
\resizebox{\hsize}{!}{\includegraphics{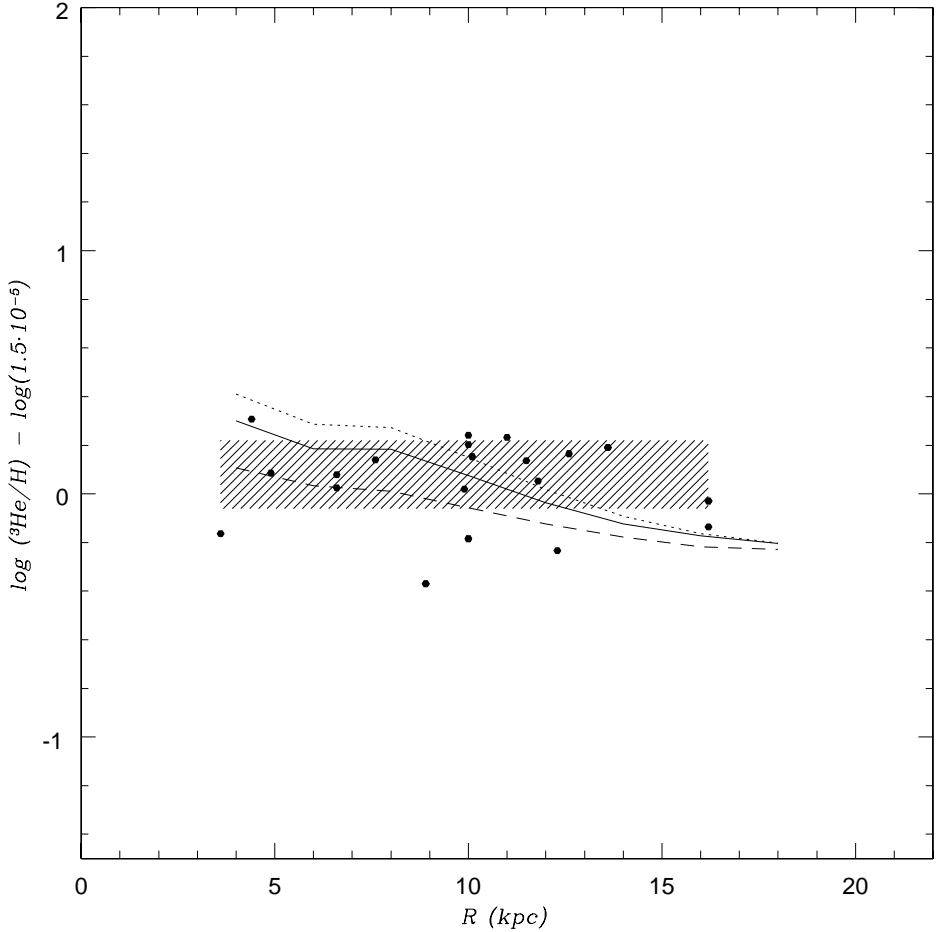}}
\caption{Distribution along the disk of $^{3}$He/H at t = 14 Gyr. Model A-I (solid), with 93\% extra-mixing, old q3 and w3 definition, ($^{3}$He/H)$_{\rm p} = 0.9\cdot 10^{-5}$; model C-I (dotted), with 93\% extra-mixing, new q3 and w3 definition, ($^{3}$He/H)$_{\rm p} = 0.9\cdot 10^{-5}$; model D-I (short-dashed), with 99\% extra-mixing, new q3 and w3 definition, ($^{3}$He/H)$_{\rm p} = 0.9\cdot 10^{-5}$. 
\newline
Observations of the hyperfine transition of $^{3}$He$^{+}$ at 8.665 GHz (3.46cm), from Galactic HII regions are shown. Abundance data are from Bania et al. (2002) (filled hexagons). The shadow area represents their
suggested average value of (1.9$\pm$0.6) $\cdot 10^{-5}$.
\newline
The predicted gradient is $-0.04<d\log$($^{3}$He/H$)$/d$R<-$0.02 dex/kpc over the 4-18 kpc range. The figure shows predictions and data normalized to the mean observed value for $^{3}$He/H in the ISM (Linsky 1998).}   
\label{Fig8}
\end{figure}

\begin{figure}
\resizebox{\hsize}{!}{\includegraphics{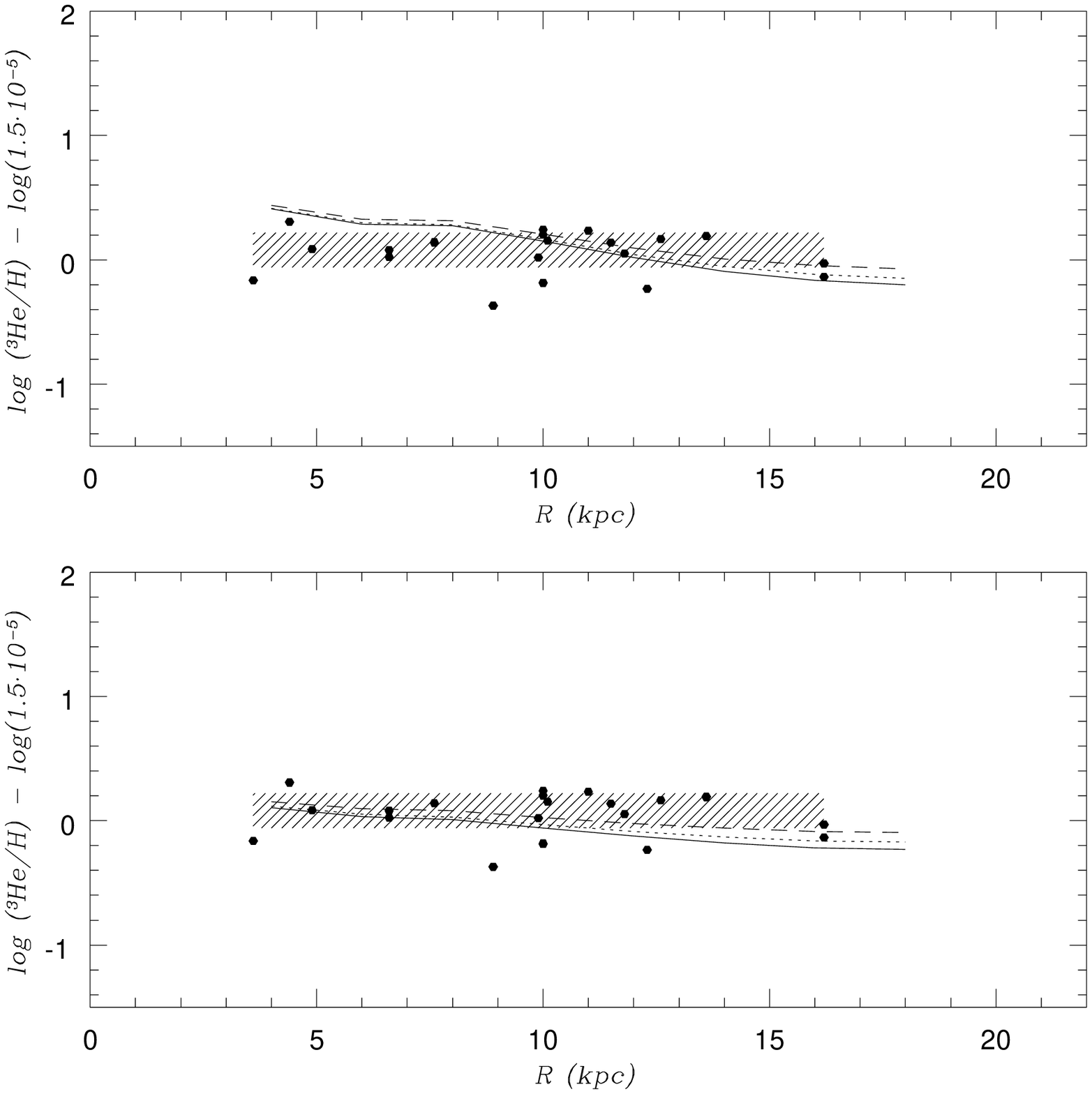}}
\caption{Predicted $^{3}$He/H gradient in the Galactic disk.
Upper panel: models C, with 93\% extra-mixing, new q3 and w3 definition -- C-I (solid), ($^{3}$He/H)$_{\rm p} = 0.9\cdot 10^{-5}$; C-II (dotted), ($^{3}$He/H)$_{\rm p} = 1.0\cdot 10^{-5}$; C-IV (short-dashed), ($^{3}$He/H)$_{\rm p} = 1.2\cdot 10^{-5}$. The predicted gradient is $-0.04<dlog^{3}He/dR<-0.03$ dex/kpc over the 4-18 kpc range.
\newline
Lower panel: models D, with 99\% extra-mixing, new q3 and w3 definition -- D-I (solid), ($^{3}$He/H)$_{\rm p} = 0.9\cdot 10^{-5}$; D-II (dotted), ($^{3}$He/H)$_{\rm p} = 1.0\cdot 10^{-5}$; D-IV (short-dashed), ($^{3}$He/H)$_{\rm p} = 1.2\cdot 10^{-5}$. With the same extra-mixing, the model using q3 and w3 redefined shows a distribution along the disk shifted toward higher values. The predicted gradient is $-0.02<d\log$($^{3}$He/H)$/d$R$<-0.01$ dex/kpc over the 4-18 kpc range. The figures show predictions and data normalized to the mean observed value for $^{3}$He/H in the ISM (Linsky 1998). Data are from Bania et al. (2002) (see Fig.8).}
\label{Fig9}
\end{figure}

\begin{figure}
\resizebox{\hsize}{!}{\includegraphics{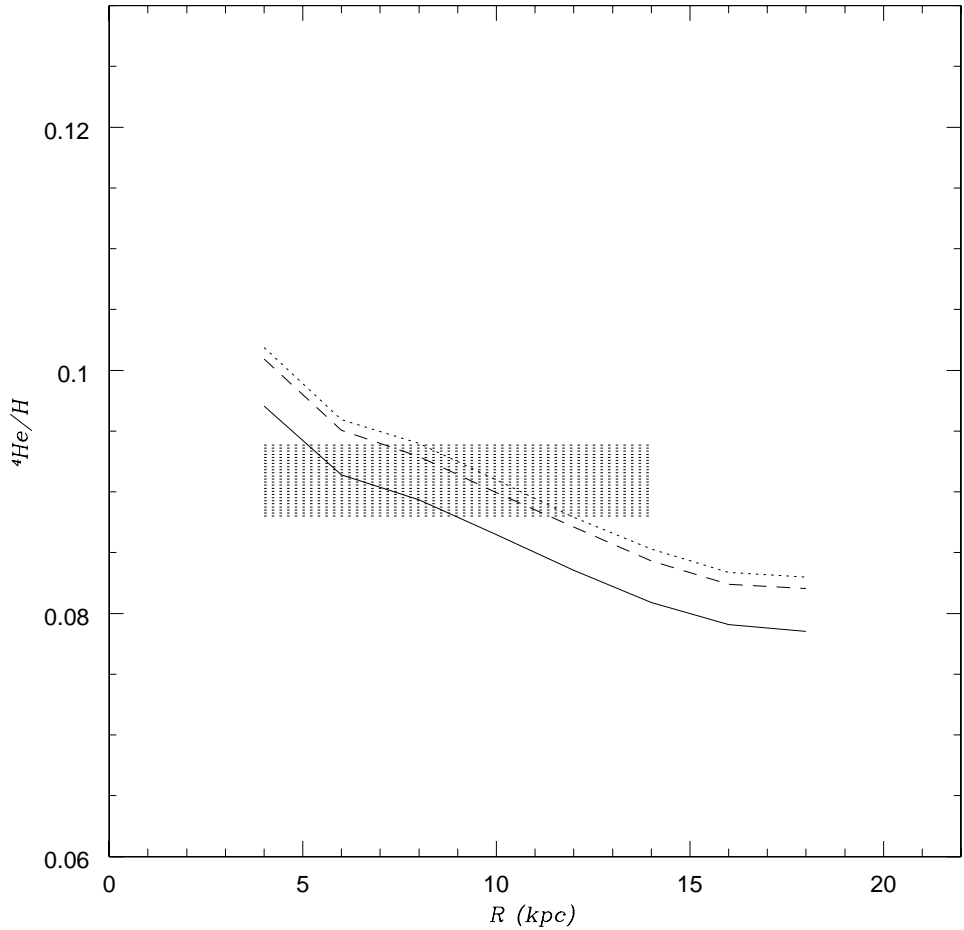}}
\caption{$^{4}$He as a function of the Galactocentric distance for models C with 93\% extra-mixing, new q3 and w3 definition: C-I (solid), $Y_{\rm p} = 0.238$; C-II (dotted), $Y_{\rm p} = 0.248$; C-IV (short-dashed), $Y_{\rm p} = 0.246$. The predicted gradient is $d$($^{4}$He/H)$/d$R$\approx-$0.002 dex/kpc over the 4-18 kpc range. The shaded region shows the locus of the planetary nebulae data of Maciel (2001).}
\label{Fig10}
\end{figure}
\newpage

\end{document}